\documentclass[10pt,twocolumn,letterpaper]{article}

\usepackage[pagenumbers]{arxiv} 
\usepackage[dvipsnames]{xcolor}
\usepackage{amssymb}

\usepackage{amsmath,amsfonts,bm}

 \def\eqref#1{(\ref{#1})}

\def\1{\bm{1}}

\DeclareMathAlphabet{\mathsfit}{\encodingdefault}{\sfdefault}{m}{sl}
\SetMathAlphabet{\mathsfit}{bold}{\encodingdefault}{\sfdefault}{bx}{n}

\usepackage{algorithm}
\usepackage{algorithmic}

\usepackage{multirow}
\usepackage{graphicx}

\definecolor{cvprblue}{rgb}{0.21,0.49,0.74}
\usepackage[pagebackref,breaklinks,colorlinks,citecolor=cvprblue]{hyperref}

\usepackage{colortbl}

\begin{document}

\title{Mitigating Data Consistency Induced Discrepancy in Cascaded Diffusion Models for Sparse-view CT Reconstruction}

\author{Hanyu Chen$^{1}$
\and Zhixiu Hao$^{1}$
\and Lin Guo$^{2}$
\and Liying Xiao$^{1*}$\\
}

\date{{$^1$Tsinghua University  \qquad $^2$Army Medical University}\\
{\tt\small xiaoly@mail.tsinghua.edu.cn}
}

\twocolumn
[{%
\renewcommand\twocolumn[1][]{#1}%
\maketitle

}]

\begin{abstract}
Sparse-view Computed Tomography (CT) image reconstruction is a promising approach to reduce radiation exposure, but it inevitably leads to image degradation. Although diffusion model-based approaches are computationally expensive and suffer from the training-sampling discrepancy, they provide a potential solution to the problem. This study introduces a novel Cascaded Diffusion with Discrepancy Mitigation (CDDM) framework, including the low-quality image generation in latent space and the high-quality image generation in pixel space which contains data consistency and discrepancy mitigation in a one-step reconstruction process. The cascaded framework minimizes computational costs by moving some inference steps from pixel space to latent space. The discrepancy mitigation technique addresses the training-sampling gap induced by data consistency, ensuring the data distribution is close to the original manifold. A specialized Alternating Direction Method of Multipliers (ADMM) is employed to process image gradients in separate directions, offering a more targeted approach to regularization. Experimental results across two datasets demonstrate CDDM's superior performance in high-quality image generation with clearer boundaries compared to existing methods, highlighting the framework's computational efficiency.
\end{abstract}

\section{Introduction}
\label{sec:intro}
Reducing the number of measurements is a commonly used method to reduce radiation exposure in X-ray Computed Tomography (CT)~\cite{brahme2014comprehensive, ravishankar2019image}. However, the resulting image quality suffers because of the ill-posed nature of the inverse problem. To address this issue, various regularization methods have been proposed~\cite{sidky2008image, bian2010evaluation, kudo2013image, hu2017improved}, but these methods are limited by the requirement for the number of projection views. The image quality degrades with artifacts when the number of views is particularly small, for instance, 16 or 8 views, because these iterative algorithms that only start from the current data are struggling with the lack of information.

With the advancement of deep learning, many data-driven models can learn the prior knowledge of CT. Models established in either the sinogram domain or the image domain can effectively enhance precision~\cite{han2018framing, zhang2018sparse, lee2018deep, liao2018adversarial, huang2020cagan}. Moreover, deep learning models can be integrated with iterative methods. These models learn priors from real images and become a part of the iterative algorithm~\cite{jin2017deep, he2018optimizing}. For example, LEARN unfolds the iterative procedure into a recurrent residual network and utilizes an adaptive regularization term with convolution neural network (CNN)~\cite{chen2018learn}. ADMM-Net defines a data-flow graph based on the Alternating Direction Method of Multipliers (ADMM) algorithm to reconstruct images, which replaces the optimization model with trained parameters~\cite{sun2016deep}.

Denoising diffusion probabilistic models (DDPM), renowned for their high-quality image generation~\cite{ho2020denoising, song2019generative, song2020score}, have become pivotal in conditional image generation tasks~\cite{dhariwal2021diffusion, nichol2021glide, ho2022classifier, saharia2022photorealistic, rombach2022high}. These models are particularly effective in both natural and medical image domains. They integrate a data fidelity term into classifier-free guidance, altering the score~\cite{kadkhodaie2021stochastic, song2021solving, chung2022improving, chung2022diffusion, he2023iterative, rout2023solving}, or use the denoised image at each diffusion step as a starting point for iterative methods~\cite{wang2022zero, chung2023solving, liu2023dolce, chung2023fast, wu2023wavelet}. However, integrating data consistency into this process presents a challenge: it can pull the denoised image away from its typical data distribution in the diffusion process. DDNM refines only the null-space contents during the reverse diffusion process by range-null space decomposition to balance the data consistency and realness~\cite{wang2022zero}, but it does not apply to the complex medical images~\cite{chung2023fast}. DiffusionMBIR chose to impose the addition total variation (TV) prior only in the redundant $z$-direction while leaving $xy$-plane intact because the diffusion prior would be corrupted by the consistency step~\cite{chung2023solving}. Our study aims to address the gap between the diffusion process and data consistency by proposing a novel approach that navigates the discrepancy between training and sampling in conditional diffusion models.

The discrepancy between training and sampling of unconditional diffusion models was first analyzed by DDPM-IP~\cite{ning2023input}, which was alleviated by input perturbation. Choosing optimal inference steps~\cite{li2023alleviating}, controlling initial signals~\cite{everaert2024exploiting}, and setting weighting strategy~\cite{yu2023debias} are some effective ways to tackle exposure bias of unconditional diffusion process. DREAM adjusts training to reflect the sampling process, which solves the discrepancy of diffusion process for image super-resolution~\cite{zhou2023dream}. Compared to the aforementioned issues, employing diffusion models for image reconstruction evidently introduces a greater discrepancy. However, there is currently a lack of methods to address this problem from the viewpoint of training diffusion models.

Inference time is also a critical obstacle for diffusion models. There are three methods to reduce the computational cost of the sampling procedure. The first is to design a more reasonable solver to reduce the number of inference steps, such as denoising diffusion implicit models (DDIM)~\cite{song2020denoising} and DPM-Solver++~\cite{lu2022dpm}. For inverse problems, reducing the number of inference steps also needs to be combined with matrix decomposition to ensure data consistency~\cite{wang2022zero, chung2023fast}. The second is to compress the image into the latent space to reduce computational costs~\cite{rombach2022high, podell2023sdxl}. However, as an image compressor, variational autoencoder (VAE) inevitably affects the quality of the image, and its image generation quality is the upper limit of image reconstruction~\cite{he2023iterative, song2023solving}. Another approach is to start the reverse diffusion process from a noisy input instead of Gaussian noise~\cite{meng2021sdedit, chung2022come, gungor2023adaptive}, and the time and initial image have a large impact on reconstruction quality.

To overcome the discrepancy introduced by data consistency and to balance the computational cost and the quality of the generated diffusion images, we propose the Cascaded Diffusion with Discrepancy Mitigation (CDDM). It is a two-stage diffusion model for sparse-view CT reconstruction, which generates low-quality images in latent space and high-quality images in pixel space. The generation process in latent space is conditioned on the image from iterative reconstruction, and random noise is added to the output image in pixel space for high-precision reconstruction. The diffusion process in pixel space is combined with data consistency and discrepancy mitigation in a one-step reconstruction procedure. Data consistency is processed by a specialized ADMM, which treats the gradients in different directions separately. Discrepancy mitigation is performed by another diffusion model to correct the error induced by data consistency and thus mitigate the discrepancy between training and sampling.

Compared to existing diffusion model-based methods to solve the inverse problem, CDDM has several innovations. First, the cascaded framework reduces the computational load by transferring some inference steps from pixel space to latent space. Second, previous studies mainly focused on designing the data consistency approach to balance the realness of solving inverse problems with diffusion models. The discrepancy between training and sampling in conditional diffusion models, which is caused by data consistency but breaks realness, has not been mentioned in previous studies. We propose a clever training objective to allow another diffusion model to learn the discrepancy induced by this guidance. During the inference process, we perform discrepancy mitigation on the image to make it closer to clean data. Third, specialized ADMM is proposed to process image gradients in separate directions, decomposing the problem in a flexible way. Our experimental results demonstrate the effectiveness of CDDM in reducing the bias for diffusion models induced by data consistency and improving the quality of the generated diffusion images.

The remainder of this paper is organized as follows. Section II provides background information on sparse-view CT reconstruction and diffusion models. Section III introduces the CDDM framework, the specialized ADMM, and discrepancy mitigation. Section IV presents the experimental results and analysis. The last section discusses related work and concludes the paper.

\section{Background}
\subsection{Sparse-view CT Reconstruction with Standard ADMM}
CT imaging, as a linear inverse problem, can be written as
\begin{equation}
\small
    {\bf{y}} = {\bf{Ax}}
\label{eqyax}
\end{equation}
where ${\mathbf{A}} \in {\mathbb{R}^{m \times n}}$ is the Radon transform, ${\mathbf{x}} \in {\mathbb{R}^n}$ is the desired image, and ${\mathbf{y}} \in {\mathbb{R}^m}$ is the measurement sinogram. The problem is ill-posed due to the sparse measured angles, which means $m \ll n$. To obtain stable solutions, it is common to transform the problem into a well-posed one by adding a regularization parameter:
\begin{equation}
\small
    {\mathbf{x}} = \mathop {\arg \min }\limits_{\mathbf{x}} \frac{1}{2}\left\| {{\mathbf{y}} - {\mathbf{Ax}}} \right\|_2^2 + \lambda R\left( {\mathbf{x}} \right)
\label{eqvarition_form}
\end{equation}
where regularization function $R\left( {\mathbf{x}} \right)$ penalizes solutions with undesired structures~\cite{benning2018modern}. TV penalty is a commonly used method for CT reconstruction, which can be expressed for 3D images as follows:
\begin{align}
\small
TV\left( {\bf{x}} \right) = {\left\| {{\bf{Dx}}} \right\|_1} \notag \\
= \sum\limits_{i,j,k} \Bigg(&\left| {{{\bf{x}}_{i,j,k}} - {{\bf{x}}_{i + 1,j,k}}} \right| + \left| {{{\bf{x}}_{i,j,k}} - {{\bf{x}}_{i,j + 1,k}}} \right| \notag \\
&+ \left| {{{\bf{x}}_{i,j,k}} - {{\bf{x}}_{i,j,k + 1}}} \right|\Bigg)
\label{eqtvdef}
\end{align}
where $\bf{D}$ is the difference matrix.

When employing the TV penalty, CT reconstruction is a Generalized Lasso problem~\cite{boyd2011distributed, chung2023solving}. The standard ADMM~\cite{ramdas2016fast} considers gradients in all directions simultaneously, whereas the specialized ADMM decomposes the difference matrix. For standard ADMM, the problem can be solved through the following algorithm:
\begin{align}
\small
{{\bf{x}}^{k + 1}} = {\left( {{{\bf{A}}^T}{\bf{A}} + \rho {{\bf{D}}^T}{\bf{D}}} \right)^{ - 1}}\left[ {{{\bf{A}}^T}{\bf{y}} + \rho {{\bf{D}}^T}\left( {{{\bf{z}}^k} - {{\bf{u}}^k}} \right)} \right]
\label{eqstandardadmmX}
\end{align}
\begin{align}
\small
{{\bf{z}}^{k + 1}} = {S_{{\lambda  \mathord{\left/
    {\vphantom {\lambda  \rho }} \right.
    \kern-\nulldelimiterspace} \rho }}}\left( {{\bf{D}}{{\bf{x}}^{k + 1}} + {{\bf{u}}^k}} \right) 
    \label{eqstandardadmmZ} 
\end{align}
\begin{align}
\small
{{\bf{u}}^{k + 1}} = {{\bf{u}}^k} + {\bf{D}}{{\bf{x}}^{k + 1}} - {{\bf{z}}^{k + 1}}
    \label{eqstandardadmmU} 
\end{align}
where ${S_\kappa }\left( a \right) = {\left( {a - \kappa } \right)_ + } - {\left( { - a - \kappa } \right)_ + }$ is the soft thresholding operator.

\subsection{Diffusion Models}
\subsubsection{DDPM}
DDPM~\cite{ho2020denoising} is a generative model to gradually add Gaussian noise to data ${{\mathbf{x}}_0} \in {\mathbb{R}^D}$, thereby forming a series of data distributions, from the real data distribution ${q_0}\left( {{{\mathbf{x}}_0}} \right)$ at time $0$ to the normal distribution ${q_T}\left( {{{\bf{x}}_T}} \right) \sim {{\cal N}}\left( {{\bf{0}},{\bf{I}}} \right)$ at time $T$. The transition distribution at time $t \in \left[ {0,T} \right]$ is 
\begin{equation}
\small
q\left( {{{\bf{x}}_t}|{{\bf{x}}_0}} \right) = {{\cal N}}\left( {{{\bf{x}}_t};\sqrt {{{\bar \alpha }_t}} {{\bf{x}}_0},\left( {1 - {{\bar \alpha }_t}} \right){\bf{I}}} \right)
\label{eqddpmtrans}
\end{equation}
where ${{\bar \alpha }_t} = \prod\nolimits_{i = 1}^t {{\alpha _i}} $, ${\alpha _t} = 1 - {\beta _t}$, and $\left\{ {{\beta _t} \in \left( {0,1} \right)} \right\}_{t = 1}^T$ is the variance scheduler. The noisy data $\mathbf{x}_t$ at arbitrary timestep $t$ can be calculated as
\begin{equation}
\small
{{\mathbf{x}}_t} = \sqrt {{{\bar \alpha }_t}} {{\mathbf{x}}_0} + \sqrt {1 - {{\bar \alpha }_t}} {{\epsilon}_t}, {{\epsilon}_t} \sim {{\cal N}}({\bf{0}},{\bf{I}}).
\label{eqDDPMtrans1}
\end{equation}

The image generation of DDPM starts from Gaussian noise ${{\bf{x}}_T}$, and gradually denoises to restore the image from the reverse diffusion process as follows:
\begin{equation}
\small
    p_{\theta} \left( \mathbf{x}_{t-1} \mid \mathbf{x}_t \right) = \cal{N} (\bf{x}_{t-1}; {\mu}_{\theta}(\bf{x}_t,t), \sigma_{t}^2\bf{I})
\label{eqDDPMreverse}
\end{equation}
where $\sigma_{t}^2$ is the untrained time-dependent constant, and the mean ${\mu}_{\theta}(\bf{x}_t,t)$ is predicted indirectly by the network whose output is the noise prediction ${{\epsilon}_\theta }\left( {{{\mathbf{x}}_t}, t} \right)$:
\begin{equation}
\small
{\mu}_{\theta}(\bf{x}_t,t) = \frac{1}{{\sqrt {{\alpha _t}} }} (\bf{x}_t - \frac{{{\beta _t}}}{{\sqrt {1 - {{\bar \alpha }_t}} }} {{\epsilon}_\theta }\left( {{{\mathbf{x}}_t}, t} \right) ).
\label{eqDDPMonestepmu}
\end{equation}
The network’s training objective is
\begin{equation}
\small
{\mathcal{L}}\left( \theta  \right) = {\mathbb{E}_{{{\mathbf{x}}_0},{{\epsilon}_t},t}}\left\| {{{\epsilon}_t} - {{\epsilon}_\theta }\left( {{{\mathbf{x}}_t}, t} \right)} \right\|^2
\label{eqDDPMobjective}
\end{equation}
where $\bf{x}_0 \sim {q_0}\left( {{{\mathbf{x}}_0}} \right)$, $\bf{\epsilon}_t \sim {{\cal N}}({\bf{0}},{\bf{I}})$ and $t \sim \mathcal{U} ([0,1, \cdots, T])$.

\subsubsection{DDIM Sampling}
DDIM~\cite{song2020denoising} achieves a faster generative process by defining a new non-Markovian diffusion process without altering the DDPM training procedure. The sampling rule can be written as
\begin{equation}
\small
    \bf{x}_{t-1} = \sqrt{{\bar \alpha}_{t-1}} {\Tilde{\mathbf{x}}_0} + \sqrt{1 - {\bar \alpha}_{t-1} - \sigma_t^2} {{\epsilon}_\theta }\left( {{{\mathbf{x}}_t}, t} \right) + \sigma_t \epsilon_{t}
\label{eqDDIMsampling}
\end{equation}
where $\Tilde{\mathbf{x}}_0$ is the predicted $\mathbf{x}_0$
\begin{equation}
\small
    \Tilde{\mathbf{x}}_0 = \frac{1}{{\sqrt {{{\bar \alpha }_t}} }}\left( {{{\mathbf{x}}_t} - \sqrt {1 - {{\bar \alpha }_t}} {{\epsilon}_\theta }\left( {{{\mathbf{x}}_t}, t} \right)} \right).
\label{eqDDIMpredictx0}
\end{equation}
The second term of Eq.~\ref{eqDDIMsampling} represents the noise addition process from clean data to noisy data, and the third term is random noise. When $\sigma_t = 0$, the image generation becomes a fixed procedure, which is the most commonly used DDIM.

\subsubsection{Latent Diffusion Model}
To reduce the computational resource consumption, latent diffusion models apply the diffusion process in the latent space of pretrained autoencoders~\cite{rombach2022high}. The encoder $\mathcal{E}$ transfers the images from pixel space $\bf{x}$ into latent space $\bf{l}$ by $\bf{l} = \mathcal{E} (\bf{x})$, and the decoder $\mathcal{D}$ reconstructs the image, giving $\mathbf{x} \approx \Tilde{\bf{x}} = \mathcal{D}(\bf{l}) = \mathcal{D}(\mathcal{E}(\bf{x}))$. The corresponding objective can be written as
\begin{equation}
\small
    {\mathcal{L}}\left( \theta  \right) = {\mathbb{E}_{{{\mathbf{l}}_0},{{\epsilon}_t},t}}\left\| {{{\epsilon}_t} - {{\epsilon}_\theta }\left( {{{\mathbf{l}}_t}, t} \right)} \right\|^2
\end{equation}
where ${\mathbf{l}_t}$ can be obtained from $\mathcal{E}$ during training.

\section{Method}

\subsection{CDDM Framework}

\begin{figure*}
    \centering
\includegraphics[width=.98\textwidth]{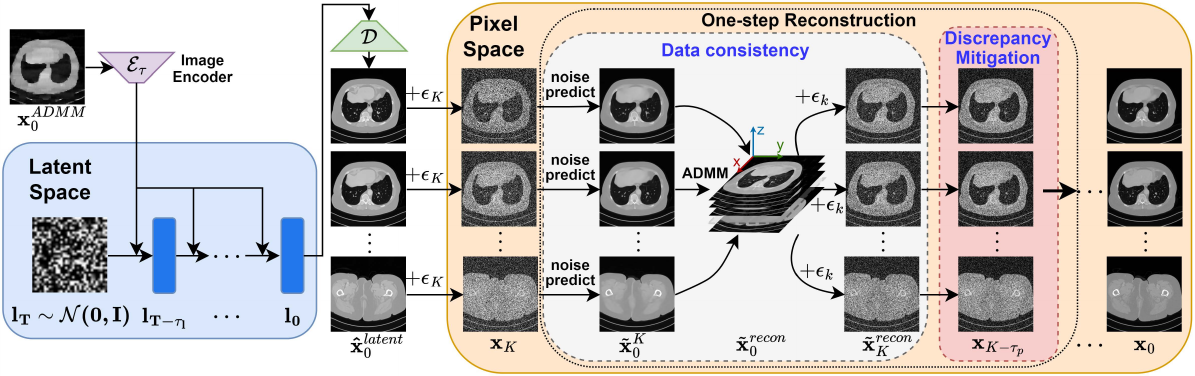}
    \caption{Overview of CDDM framework for sparse-view CT reconstruction. A standard diffusion process in latent space generates low-quality CT images, with the image condition provided by specialized ADMM. Noise is then added to the images in pixel space and the diffusion process is utilized again to generate high-quality CT images. Specialized ADMM treats the gradients of different directions separately, and the discrepancy mitigation uses another diffusion process to correct errors induced by data consistency.
    }
    \label{fig:overview}
    \vspace{-.1in}
\end{figure*}

The framework of the CDDM is depicted in Fig.~\ref{fig:overview}, which consists of a latent diffusion model and a pixel diffusion model connected in series. The sinogram $\mathbf{y}$ is reconstructed by a specialized ADMM with initial point $\mathbf{A}^T \mathbf{y}$. The reconstructed image $\mathbf{x}_0^{ADMM}$ is then extracted by image encoder $\cal{E}_{\tau}$ and is input as the condition to the latent diffusion network $\theta_l$, which is fine-tuned from Stable Diffusion~\cite{rombach2022high}. The low-quality image $\mathbf{\hat x}_0^{latent} = {\cal{D}}(\mathbf{l}_{0})$ is generated from noise $\mathbf{l}_T \sim \cal{N} {(\mathbf{0}, \mathbf{I})}$ using the fast sampler DDIM to accelerate the generation processes.

The high-quality image is then generated in pixel space from corrupted data $\mathbf{x}_K$ using Eq.~\ref{eqDDPMtrans1} with $K = t_0 T$, where $T$ is the total inference steps in pixel diffusion and noise strength $t_0 < 1$. The one-step reconstruction step is composed of data consistency and discrepancy mitigation (DM). The specific process is as follows:
\begin{itemize}
    \item Data consistency performs the specialized ADMM optimization in the clean data manifold from the denoised $\tilde{{\mathbf{{x}}}}_0^{k}$. The regularization term of this approach is decomposed into two parts: one performed on the $xy$-plane and one on the $z$-axis, which will be further discussed in Sec.~\ref{secsADMM}. This optimization approach allows for a better combination of DM, utilizing the $z$-axis knowledge while obtaining a better $xy$-plane prior.
    \item DM reduces the error through another diffusion process. The noise is added to reconstructed $\tilde{\mathbf{x}}_0^{recon}$ to obtain $\tilde{\mathbf{x}}_k^{recon}$ rather than $\tilde{\mathbf{x}}_{k-\tau_p}^{recon}$. DM is then performed to denoise the image and to eliminate the training-sampling discrepancy induced by data consistency, which utilizes further fine-tuned diffusion network $\theta_{p^{\prime}}$ in Sec.~\ref{secDM}.
\end{itemize}
The final reconstructed image $\mathbf{x}_0$ can be obtained by iterating this step. The procedure of the CDDM algorithm is shown in Algorithm~\ref{alg:CDDM}, which outlines the inference steps involved in the CDDM framework and shows the integration of diffusion models.

\begin{algorithm}[t]
\renewcommand{\algorithmicrequire}{\textbf{Input:}}
\renewcommand{\algorithmicensure}{\textbf{Output:}}
\small
\caption{CDDM}
\begin{algorithmic}[1]\label{alg:CDDM}
\REQUIRE CLIP image encoder $\cal{E}_{\tau}$, measured sinogram $\mathbf{y}$, noise $\mathbf{l}_T$,
latent diffusion forward steps $T$ and sampling interval $\tau_l$, VAE decoder $\cal{D}$, pixel diffusion steps $K$ and sampling interval $\tau_p$
\STATE $\mathbf{x}_0^{ADMM} = ADMM(\cdot) $ 
\COMMENT {Initial point $\mathbf{A}^T\mathbf{y}$}
\FOR {$k$ in $\{ T, T-\tau_s, \cdots, \tau_s\}$} 
    \STATE $\mathbf{l}_{k-\tau_s} = DDIM(\mathbf{l}_k, \epsilon_{\theta_{l}})$
\ENDFOR
\vspace{2mm}
\STATE $\mathbf{\hat x}_0^{latent} = {\cal D} (\mathbf{l}_0) $
\STATE $\mathbf{x}_K = \sqrt{\bar \alpha_K} \mathbf{\hat x}_0^{latent} + \sqrt{1 - \bar \alpha_K} \mathbf{\epsilon}_K$
\vspace{2mm}
\FOR {$k$ in $\{ K, K-\tau_p, \cdots, \tau_p\}$} 
    \STATE $\tilde{{\mathbf{{x}}}}_0^{k} = \frac{1}{{\sqrt {{{\bar \alpha }_k}} }}\left( {{{\mathbf{x}}_k} - \sqrt {1 - {{\bar \alpha }_k}} {{\epsilon}_{\theta_{p}} }\left( {{{\mathbf{x}}_k}, k} \right)} \right)$
    \STATE $\tilde{\mathbf{x}}_0^{recon} = ADMM(\cdot)$
    \COMMENT{Initial point $\tilde{{\mathbf{{x}}}}_0^{k}$}
    \STATE $\tilde{\mathbf{x}}_k^{recon} = \sqrt{\bar \alpha_k} \tilde{\mathbf{x}}_0^{recon} + \sqrt{1 - \bar \alpha_k}  {{\epsilon}_{\theta_{p}} }\left( {{{\mathbf{x}}_k}, k} \right)$
    \STATE $\mathbf{x}_{k-\tau_p} = DDIM(\tilde{\mathbf{x}}_k^{recon}, \epsilon_{\theta_{p^{\prime}}})$
    \COMMENT{Discrepancy Mitigation}
\ENDFOR
\ENSURE $\mathbf{x}_0$
\end{algorithmic}
\end{algorithm}

\subsection{Specialized ADMM for Inverse Problem}
\label{secsADMM}
CDDM processes the denoised CT image of every step with data consistency using the ADMM method, which necessitates the consideration of gradients in three directions. Previous studies utilize the standard ADMM to solve the TV regularization problem, which treats the gradients in all directions as a whole. In our study, we propose a specialized ADMM to process the gradients in different directions separately, with $\mathbf{D} = \mathbf{D}_x  \mathbf{D}_y \mathbf{D}_z = \mathbf{D}_{xy}  \mathbf{D}_z$. The specialized approach decomposes a difficult problem into two relatively complex subproblems, as opposed to one complex and one simple subproblem~\cite{ramdas2016fast, barbero2014modular}. Existing studies of specialized ADMM are primarily focused on filter trending or image denoising problems, and do not tackle the more complex inverse problem where the forward operator $\mathbf{A} \neq \mathbf{I}$. Therefore, we discuss CT reconstruction using specialized ADMM in detail in this part.

Combining the $xy$ terms in the regularization term of Eq.~\ref{eqvarition_form}, the 3D fused LASSO problem can be written as:
\begin{align}
\small
    {\mathbf{x}} = \mathop {\arg \min }\limits_{\mathbf{x}} \frac{1}{2}\left\| {{\mathbf{y}} - {\mathbf{Ax}}} \right\|_2^2 + \lambda ({\left\| {{\mathbf{D}_{xy} \bf{x}}} \right\|_1}+ {\left\| {{\mathbf{D}_z \bf{x}}} \right\|_1}).
\label{eq3DfusedLASSO}
\end{align}
Rewrite the problem to separate the variables:
\begin{align}
\small
    {\mathbf{h}},{\mathbf{v}} = \mathop {\arg \min }\limits_{\mathbf{h},{\mathbf{v}}} \frac{1}{2} \left\| {{\mathbf{y}} - {\mathbf{Ah}}} \right\|_2^2 &+ \lambda ({\left\| {{\mathbf{D}_{xy} \bf{h}}} \right\|_1}+ {\left\| {{\mathbf{D}_z \bf{v}}} \right\|_1}) \notag \\
    &s.t. \quad {\mathbf{h}} - {\mathbf{v}} = 0,
\end{align}
where $\mathbf{h}$ and $\mathbf{v}$ are both 3D CT images but they are used to penalize the gradients in various directions. Then we yield the ADMM updates:
\begin{align}
\small
        {\mathbf{h}^{k+1} } = \mathop {\arg \min }\limits_{\bf{h}}  &{\frac{1}{2}\left\| {\bf{y} - \bf{Ah}} \right\|_2^2 + \lambda  {\left\| {{\mathbf{D}_{xy} \bf{h}}} \right\|_1  } }\notag \\
    &+ \frac{{{\rho _1}}}{2}\left\| {\mathbf{h} - {\mathbf{v}}^{k} + {\mathbf{w}}^{k}} \right\|_2^2 
\label{eqSpecializedADMMH}
\end{align}
\vspace{-7ex}

\begin{align}
\small
    {\mathbf{v}^{k+1} } = \mathop {\arg \min }\limits_{\bf{v}}  \lambda  {\left\| {{{\mathbf{D}}_z \bf{v}}} \right\|_1  } + \frac{{{\rho _1}}}{2}\left\| {\mathbf{h}^{k+1} - \mathbf{v} + \mathbf{w}^k} \right\|_2^2
\label{eqSpecializedADMMV}
\end{align}
\vspace{-7ex}

\begin{align}
\small
    {\mathbf{w}^{k+1} } = \mathbf{w}^k +\mathbf{h}^{k+1} - \mathbf{v}^{k+1}.
\label{eqSpecializedADMMW}
\end{align}
The subproblems of Eq.~\ref{eqSpecializedADMMH} and Eq.~\ref{eqSpecializedADMMV} can still be solved by ADMM with closed form solution:
\begin{flalign}
\small
    &\ \hspace{5mm}  {{\mathbf{h}}^{k+1} } = \mathbf{A}_h^{ - 1} \mathbf{b}_{h} &
\label{eqADMMsubproblemH}
\end{flalign}
\vspace{-7ex}

\begin{flalign}
\small
    &\ \hspace{5mm} \mathbf{z}_{h}^{k+1} = {S_{{\lambda  \mathord{\left/
    {\vphantom {\lambda  {{\rho _2}}}} \right.
    \kern-\nulldelimiterspace} {{\rho _2}}}}} (\mathbf{D}_{xy} \mathbf{h}^{k+1} + \mathbf{s}_{h}^{k}) &
\label{eqADMMsubproblemZH}
\end{flalign}
\vspace{-7ex}

\begin{flalign}
\small
    &\ \hspace{5mm} \mathbf{s}_{h}^{k+1} = \mathbf{s}_{h}^{k} + \mathbf{D}_{xy} \mathbf{h}^{k+1} - \mathbf{z}_{h}^{k+1} &
\label{eqADMMsubproblemSH}
\end{flalign}
\vspace{-7ex}

\begin{flalign}
\small
    &\ \hspace{5mm} {{\mathbf{v}}^{k+1} } = \mathbf{A}_v^{ - 1} \mathbf{b}_{v} &
\label{eqADMMsubproblemV}
\end{flalign}
\vspace{-7ex}

\begin{flalign}
\small
    &\ \hspace{5mm} \mathbf{z}_{v}^{k+1} = {S_{{\lambda  \mathord{\left/
    {\vphantom {\lambda  {{\rho _1 \rho_3}}}} \right.
    \kern-\nulldelimiterspace} {{\rho _1 \rho_3}}}}} (\mathbf{D}_{z} \mathbf{v}^{k+1} + \mathbf{S}_{v}^{k}) &
\label{eqADMMsubproblemZV}
\end{flalign}
\vspace{-7ex}

\begin{flalign}
\small
    &\ \hspace{5mm} \mathbf{s}_{v}^{k+1} = \mathbf{s}_{v}^{k} + \mathbf{D}_{z} \mathbf{v}^{k+1} - \mathbf{z}_{v}^{k+1} &
\label{eqADMMsubproblemSV}
\end{flalign}
where the specific formulation of Eq.~\ref{eqADMMsubproblemH} can be written as $\left\{ {\begin{array}{*{20}{l}}
    \mathbf{A}_{h} =  {{{\mathbf{A}}^T}{\mathbf{A}} + {\rho _1}{\mathbf{I}} + {\rho _2}{{\mathbf{D}}_{xy}}^T{{\mathbf{D}}_{xy}}} \\ 
    \mathbf{b}_{h} = {{{\mathbf{A}}^T}{\mathbf{y}} - {\rho _1}{{\left( {{\mathbf{w}^k} - {\mathbf{v}^k}} \right)}^T} + {\rho _2}{{\mathbf{D}}_{xy}}^T\left( {{\mathbf{z}_{h}^{k}} - {\mathbf{s}_{h}^{k}}} \right)}
\end{array}} \right.$ and the specific formulation of Eq.~\ref{eqADMMsubproblemV} can be written as $\left\{ {\begin{array}{*{20}{l}}
    \mathbf{A}_{v} =  {{\mathbf{I}} + {\rho _3} {{\mathbf{D}}_z}^T{{\mathbf{D}}_z}} \\ 
    \mathbf{b}_{v} = \mathbf{h}^{k+1} + \mathbf{w}^k + {\rho _3}{{\mathbf{D}}_z}^T\left( {{\mathbf{z}_{v}^{k}} - {\mathbf{s}_{v}^{k}}} \right)
\end{array}} \right.$.

Considering the complexity of matrix inversion, the conjugate gradient (CG) method is used to solve Eq.~\ref{eqADMMsubproblemH} and Eq.~\ref{eqADMMsubproblemV}~\cite{chung2023solving}, which also guarantees the updated samples stay within the tangent space~\cite{chung2023fast}. The detailed algorithm is proposed in Algorithm~\ref{alg:specializedADMM}, where the 3D CT image reconstruction problem is split into two subproblems solved by ADMM. Theoretically, each subproblem should be iterated in N steps, but we empirically let $N = 1$ to perform the reconstruction. The detailed parameter settings are presented in Table~\ref{tab_admm_paras}.

\begin{algorithm}[t]
\small
\caption{Specialized ADMM}
\begin{algorithmic}[1]\label{alg:specializedADMM}
\FOR {$k = 1$ to $K$}
    \STATE {Update $\mathbf{h}^{k+1}$, $\mathbf{z}_h^{k+1}$ and $\mathbf{s}_h^{k+1}$ with Eq.~\ref{eqADMMsubproblemH}} to Eq.~\ref{eqADMMsubproblemSH}
    \STATE {Update $\mathbf{v}^{k+1}$, $\mathbf{z}_v^{k+1}$ and $\mathbf{s}_v^{k+1}$ with Eq.~\ref{eqADMMsubproblemV}} to Eq.~\ref{eqADMMsubproblemSV}
    \STATE {${\mathbf{w}^{k+1} } = \mathbf{w}^k +\mathbf{h}^{k+1} - \mathbf{v}^{k+1}$}
\ENDFOR
\end{algorithmic}
\end{algorithm}

\subsection{Discrepancy Mitigation}
\label{secDM}
Data consistency is a crucial step in solving the inverse problem, but it may introduce errors to the image that is not consistent with the original distribution trained by the diffusion model. Therefore, discrepancy mitigation is proposed to correct the error induced by data consistency by another diffusion process. Similarly to the diffusion rectification from DREAM~\cite{zhou2023dream}, the training objective of DM is extended to align with the diffusion sampling with data consistency, but it is more complex due to the inclusion of the reconstruction step.

During training, noisy image $\bf{x}_t$ is obtained by Eq.~\ref{eqDDPMtrans1} and $\Tilde{\mathbf{x}}_0$ is predicted as normal DDIM. The predicted $\Tilde{\mathbf{x}}_0$ will be used for data consistency, just like the sampling process to get the reconstructed $\Tilde{\mathbf{x}}_0^{recon} = ADMM(\Tilde{\mathbf{x}}_0)$. Since the images during training are single-slice, the specialized ADMM does not take into account the effect of the $z$-direction gradient, and the gradients on $xy$-plane are separated during training. Then Eq.~\ref{eqSpecializedADMMH} and Eq.~\ref{eqADMMsubproblemV} becomes:
\begin{align}
\small
        {\mathbf{h}^{k+1} } = \mathop {\arg \min }\limits_{\bf{h}}  &{\frac{1}{2}\left\| {\bf{y} - \bf{Ah}} \right\|_2^2 + \lambda  {\left\| {{\mathbf{D}_{x} \bf{h}}} \right\|_1  } }\notag \\
    &+ \frac{{{\rho _1}}}{2}\left\| {\mathbf{h} - {\mathbf{v}}^{k} + {\mathbf{w}}^{k}} \right\|_2^2 
\label{eqSpecializedADMM_trainH}
\end{align}
\vspace{-5ex}
\begin{align}
\small
    {\mathbf{v}^{k+1} } = \mathop {\arg \min }\limits_{\bf{v}}  \lambda  {\left\| {{{\mathbf{D}}_y \bf{v}}} \right\|_1  } + \frac{{{\rho _1}}}{2}\left\| {\mathbf{h}^{k+1} - \mathbf{v} + \mathbf{w}^k} \right\|_2^2 
\label{eqSpecializedADMM_trainV}
\end{align}
and the update method is the same as that in Sec.~\ref{secsADMM}.

The self-estimated clean image will be added by new noise $\epsilon_t^{\prime}$ to get a new noisy image $\Tilde{\mathbf{x}}_t^{recon}$ to serve as input to the network
\begin{equation}
\small
    \Tilde{\mathbf{x}}_t^{recon} = \sqrt {{{\bar \alpha }_t}} {\Tilde{\mathbf{x}}_0^{recon}} + \sqrt {1 - {{\bar \alpha }_t}} {{\epsilon}_t^{\prime}}
\label{eqreconnoise}
\end{equation}
where ${{\epsilon}_t^{\prime}} \sim {{\cal N}}({\bf{0}},{\bf{I}})$. Then, the relationship between the reconstructed noisy image $\Tilde{\mathbf{x}}_t^{recon}$ and the original image ${\mathbf{x}}_0$ becomes
\begin{align}
\small
    \Tilde{\mathbf{x}}_t^{recon} &= \sqrt {{{\bar \alpha }_t}} {\mathbf{x}_0} + \sqrt {1 - {{\bar \alpha }_t}} {{\epsilon}_t^{\prime}} + \sqrt {{{\bar \alpha }_t}} \left({\Tilde{\mathbf{x}}_0^{recon}} - {\mathbf{x}_0} \right) \notag \\
    &= \sqrt {{{\bar \alpha }_t}} {\mathbf{x}_0} \notag \\
    &\hspace{0.2cm}+ \sqrt {1 - {{\bar \alpha }_t}} \left( {{\epsilon}_t^{\prime}} + \sqrt {\frac{{{{\bar \alpha }_t}}}{{1 - {{\bar \alpha }_t}}}}  \left({\Tilde{\mathbf{x}}_0^{recon}} - {\mathbf{x}_0} \right) \right) \notag \\
    &= \sqrt {{{\bar \alpha }_t}} {\mathbf{x}_0} + \sqrt {1 - {{\bar \alpha }_t}} \left( {{\epsilon}_t^{\prime}} + \sqrt {\frac{{{{\bar \alpha }_t}}}{{1 - {{\bar \alpha }_t}}}} \Delta \mathbf{x}_0 \right)
\label{eqreconxt}
\end{align}
where $\Delta \mathbf{x}_0 = {\Tilde{\mathbf{x}}_0^{recon}} - {\mathbf{x}_0} $ is the difference between reconstructed and original image. The training objective for this diffusion model can be expressed as
\begin{align}
\small
{\mathcal{L}}\left( \theta  \right) =  {\mathbb{E}_{{{\mathbf{x}}_0},{{\epsilon}_t^{\prime}},t}}\left\| {\left( {{\epsilon}_t^{\prime}} + \sqrt {\frac{{{{\bar \alpha }_t}}}{{1 - {{\bar \alpha }_t}}}}  \Delta \mathbf{x}_0 \right) - {{\epsilon}_\theta }\left( {\Tilde{\mathbf{x}}_t^{recon}, t} \right)} \right\|^2.
\label{eqnewobj}
\end{align}

\begin{algorithm}[t]
    \small
    \caption{DM Training}
    \begin{algorithmic}[1]\label{alg:dm-training}
        \REPEAT
        \STATE $\bf{x}_0 \sim {q_0}\left( {{{\mathbf{x}}_0}} \right)$, $\bf{\epsilon}_t \sim {{\cal N}}({\bf{0}},{\bf{I}})$, $t \sim \mathcal{U} ([0,1, \cdots, T])$
        \STATE ${{\mathbf{x}}_t} = \sqrt {{{\bar \alpha }_t}} {{\mathbf{x}}_0} + \sqrt {1 - {{\bar \alpha }_t}} {{\epsilon}_t}$
        \STATE $\Tilde{\mathbf{x}}_0 = \frac{1}{{\sqrt {{{\bar \alpha }_t}} }}\left( {{{\mathbf{x}}_t} - \sqrt {1 - {{\bar \alpha }_t}} {{\epsilon}_\theta }\left( {{{\mathbf{x}}_t}, t, c_1} \right)} \right)$
        \STATE Take gradient descent step on $\nabla \left\| {{{\epsilon}_t} - {{\epsilon}_\theta }\left( {{{\mathbf{x}}_t}, t, c_1} \right)} \right\|^2$
        \STATE $\Tilde{\mathbf{x}}_0^{recon} = ADMM(\Tilde{\mathbf{x}}_0)$
        \STATE $\Tilde{\mathbf{x}}_t^{recon} = \sqrt {{{\bar \alpha }_t}} {\Tilde{\mathbf{x}}_0^{recon}} + \sqrt {1 - {{\bar \alpha }_t}} {{\epsilon}_t^{\prime}}$
        \STATE $\Delta \mathbf{x}_0 = {\Tilde{\mathbf{x}}_0^{recon}} - {\mathbf{x}_0} $
        \STATE Take gradient descent step on \notag \\  $\nabla \left\| {\left( {{\epsilon}_t^{\prime}} + \lambda_{\alpha}  \Delta \mathbf{x}_0 \right) - {{\epsilon}_\theta }\left( {\Tilde{\mathbf{x}}_t^{recon}, t, c_2} \right)} \right\|^2$
        \UNTIL{converged}
    \end{algorithmic}
    \end{algorithm}

However, when $t \to 0$, the coefficient $\sqrt {\frac{{{{\bar \alpha }_t}}}{{1 - {{\bar \alpha }_t}}}} $ becomes very large, leading to loss exposure, which poses difficulties in network training. Therefore, it is necessary to limit the size of this coefficient with $\lambda_{max}$ and rewrite the objective Eq.~\ref{eqnewobj} as follows:
\begin{align}
\small
{\mathcal{L}}\left( \theta  \right) =  {\mathbb{E}_{{{\mathbf{x}}_0},{{\epsilon}_t^{\prime}},t}}\left\| {\left( {{\epsilon}_t^{\prime}} + \lambda_{\alpha}  \Delta \mathbf{x}_0 \right) - {{\epsilon}_\theta }\left( {\Tilde{\mathbf{x}}_t^{recon}, t} \right)} \right\|^2
\label{eqnewobjrewrite}
\end{align}
where $\lambda_{\alpha} = \mathop{\min} \left(\sqrt {\frac{{{{\bar \alpha }_t}}}{{1 - {{\bar \alpha }_t}}}}, \lambda_{max} \right)$. As the parameter $t$ increases, it follows that $\lambda_{\alpha} = \sqrt{\frac{{\bar{\alpha}_t}}{{1 - \bar{\alpha}_t}}}$, indicating that the diffusion process increasingly approximates the prediction of the original $\mathbf{x}_0$. Conversely, as $t$ approaches zero, $\lambda_{\alpha}$ is less than $\sqrt{\frac{{\bar{\alpha}_t}}{{1 - \bar{\alpha}_t}}}$, suggesting a preference for the generation of reconstructed images $\Tilde{\mathbf{x}}_t^{recon}$. This phenomenon aligns with the process of image generation, wherein at the initial stages of the process, the image is required to conform to the distribution of actual images. Conversely, towards the conclusion of the generation process (where noise levels are minimized), the image is anticipated to bear a closer resemblance to the measured data $\mathbf{y}$.

In practical applications, the DM training employs identical Gaussian noise ($\epsilon_t = \epsilon_t^{\prime}$) as delineated in DREAM~\cite{zhou2023dream}. However, it is necessary to retain the gradient from the initial prediction step according to Eq.~\ref{eqDDPMobjective} to prevent the catastrophic forgetting of authentic images. To facilitate the differentiation between the two categories of images, a label $c \in {c_1, c_2}$ is incorporated into the input of the network. Here, $c_1$ and $c_2$ denote real and reconstructed images, respectively. The detailed training algorithm is presented in Algorithm~\ref{alg:dm-training}.

\section{Experiments and Results}
\subsection{Experiment Setup}
\subsubsection{Dataset Preparation}

In our study, we utilized two distinct datasets to evaluate the effectiveness of the CDDM framework: the Walnut dataset~\cite{der2019cone} and the AAPM 2016 Low Dose CT Grand Challenge~\cite{moen2021low}. The Walnut dataset consists of high-resolution cone-beam CT scans of 42 walnuts, providing a diverse range for training ( Walnut $\#1-37$, 13822 slices) and testing (Walnut $\#38,39,40,42$, 1425 slices). The walnut $\#41$ was excluded due to its artifacts. Additionally, the reconstructed images are center cropped on the axial plane to the resolution of $256 \times 256$ and pixel size $1.5\,mm \times 1.5\,mm$. The AAPM dataset comprises 10 abdominal CT volumes from anonymous patients, which were divided into training (9 patients, 4090 slices) and testing (1 patient, 500 slices) sets. The generation of 8-view sinograms for these datasets was performed using $\textit{TorchRadon}$ library~\cite{ronchetti2020torchradon}.

\subsubsection{Implementation Details}

The networks for diffusion models were trained using the two datasets separately, and the training hyperparameters were selected on the Walnut training dataset. The latent models were finetuned from $\textit{Stable Diffusion v-1.4}$ with $50k$ training steps, batch size $16$ and learning rate $2\times10^{-5}$, where the CLIP text encoder was replaced by MedCLIP~\cite{wang2022medclip} vision decoder. The pixel diffusion models had a similar structure to the latent diffusion model, but the cross-attention layers with the encoder were removed. The networks were randomly initialized and were trained with $50k$ steps, batch size $2$, and learning rate $1\times10^{-5}$. The DM network was finetuned from the pixel diffusion model with $5k$ steps, batch size $1$, and learning rate $1\times10^{-5}$.

For CDDM inference, the specialized ADMM was first implemented to serve as the condition of latent diffusion and to guide the diffusion process as a data consistency method. Parameters were selected using a grid search method on a small CT volume of Walnut $\#1$ with dimensions $10 \times 256 \times 256$ and a CT volume of a patient in the training dataset of the same size. The details of the parameters are presented in Table~\ref{tab_admm_paras}. The latent diffusion employs the DDIM sampling method, with forward diffusion steps $T = 1000$, a sampling interval $\tau_l = 20$, and a classifier-free guidance strength $w=2$. The inference parameters of pixel diffusion demonstrated optimal performance with a noise strength of 0.5 and a sampling interval $\tau_p = 20$.

\begin{table}[t!]
    \centering
    \caption{Parameters of specialized ADMM selected by grid search.}
    \small
    \resizebox{0.82\columnwidth}{!}{
    \begin{tabular}{ccccccc}
    \toprule
        Dataset & \multicolumn{1}{|c|}{Stage} & $\rho_1$ & $\rho_2$ & $\rho_3$ & $\lambda$ & $K$\\
        \hline
        \multirow{2}{*}{Walnut} & \multicolumn{1}{|c|}{Low-quality Generation} & 2 & 0.8 & 150 & 1.2 & 50 \\
            & \multicolumn{1}{|c|}{Data Consistency} & 0.01 & 10 & 50 & 0.2 & 10\\
        \hline
        \multirow{2}{*}{AAPM} & \multicolumn{1}{|c|}{Low-quality Generation} & 1 & 0.3 & 80 & 0.8 & 100\\
            & \multicolumn{1}{|c|}{Data Consistency} & 0.01 & 10 & 100 & 0.02 & 10\\
        \bottomrule
    \end{tabular}}
    \vspace{-2em}
    \label{tab_admm_paras}
\end{table}

\subsubsection{Evaluation Methods}
Our proposed approach, CDDM, was compared with several benchmarks, including DDS~\cite{chung2023fast}, FBPConvNet~\cite{jin2017deep}, and the standard ADMM~\cite{boyd2011distributed}, covering a spectrum of commonly utilized diffusion processes, data-driven models, and iterative methodologies. DDS employs the same diffusion model as CDDM, with the data consistency parameters $(\lambda, \rho) = (0.1, 10)$ for the Walnut dataset and $(0.04, 10)$ for the AAPM dataset, respectively. The inference steps $T$ were set to 50. FBPConvNet was trained on 2D images over 100 epochs. The ADMM approach, used for comparison, was the same as the conditional images for latent diffusion. The reconstruction quality was assessed via peak signal-to-noise ratio (PSNR) and structural similarity index measure (SSIM), examining metrics across axial, coronal, and sagittal planes for each CT volume.

\subsection{Experiment results}

\begin{figure*}[t!]
    \centering
    \includegraphics[width=0.69\linewidth]{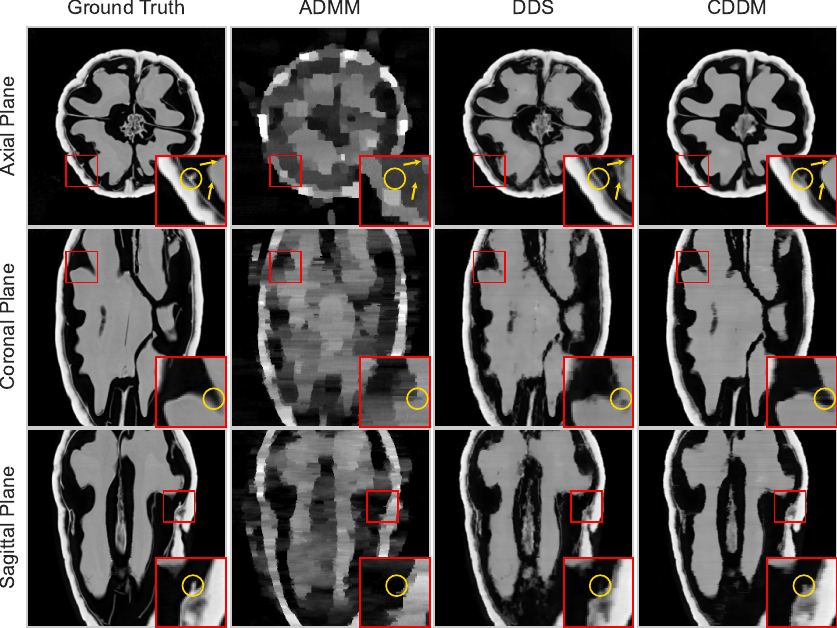}
    \vspace{-.1in}
    \caption{
    Representative results of 8-view CT reconstruction performance on Walnut dataset. From left to right, ground truth, ADMM, DDS, and our proposed CDDM. The three columns show reconstruction images of the axial, coronal, and sagittal plane from top to bottom. For ease of presentation, the rectangular coronal and sagittal images were center cropped to become squares.} 
        \label{fig_walnut}
    \vspace{-.1in}
\end{figure*}

\begin{figure}[!ht]
    \centering
    \includegraphics[width=0.99\columnwidth]{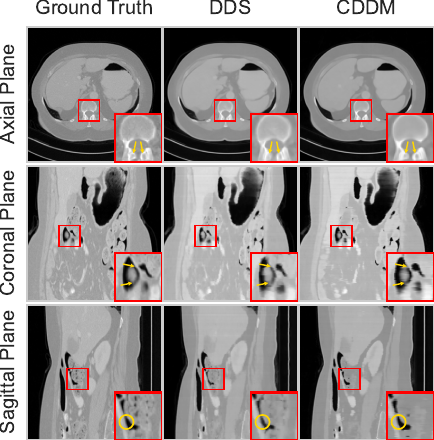}
    \vspace{-.1in}
    \caption{
    Representative results of 8-view CT reconstruction performance on AAPM dataset. From left to right, ground truth, DDS, and our proposed CDDM. The three columns show reconstruction images of the axial, coronal, and sagittal plane from top to bottom.} 
        \label{fig_aapm}
    \vspace{-.1in}
\end{figure}

Table~\ref{tab_res_walnuts} and Table~\ref{tab_res_aapm} present the quantitative results of various reconstruction methods on the Walnut and AAPM datasets, respectively. According to the outcomes, for the ultra-sparse view reconstruction challenge, neither the iterative method ADMM nor the end-to-end deep learning model FBPConvNet achieves satisfactory results. Methods based on diffusion generative models yield favorable outcomes. On the Walnut dataset, CDDM surpasses DDS in performance across all planes. Their SSIM metrics are comparable, indicating a general closeness in structural information. However, CDDM's superior PSNR metrics across all directions suggest enhanced local detail and texture reconstruction, as well as more effective noise reduction. Similar trends are observed in the AAPM dataset results, with CDDM achieving superior PSNR in all planes, albeit slightly lower in SSIM on the coronal and sagittal planes. Compared between the two datasets, metrics are generally lower for the Walnut dataset due to its fewer homogeneous regions and more intricate structures than the human data of the AAPM dataset.
Fig.~\ref{fig_walnut} presents a comparison of the reconstruction images from ADMM, DDS, and CDDM on the Walnut dataset. The results clearly show that the iterative method ADMM struggled to reconstruct the images effectively, with poor boundary definition and a lack of detail. As seen in the region of interest (ROI) marked with green circles and arrows, CDDM outperforms DDS in terms of image quality, with clearer boundaries to distinguish the walnut shell and more detailed structures in axial, coronal, and sagittal planes. The feature of the CDDM results on the axial plane suggests that DM is effective in reducing the bias from data consistency and enhancing image quality. The images on the coronal and sagittal planes demonstrate that specialized ADMM is capable of guiding the diffusion process to achieve better performance by considering the gradient terms. Fig.~\ref{fig_aapm} shows the comparison of the reconstruction images from DDS and CDDM on the AAPM dataset. The results also illustrate that CDDM obtains better image quality than DDS. The marked arrow in the ROI presents image edges to determine the human body structure. DDS results can show unclear boundaries of body tissues, which can interfere with diagnosis, while CDDM image results have clearer edges.
\begin{table}[t!]
    \centering
    \caption{Quantitative results from different reconstruction algorithms for Walnut dataset.}
    \small
    \resizebox{\columnwidth}{!}{
    \begin{tabular}{lcccccc}
    \toprule
            \multicolumn{1}{c|}{} & \multicolumn{2}{c|}{Axial Plane} & \multicolumn{2}{c|}{Coronal Plane} & \multicolumn{2}{c}{Sagittal Plane}\\
        \multicolumn{1}{c|}{} & PSNR$\uparrow$ & \multicolumn{1}{c|}{SSIM$\uparrow$} & PSNR$\uparrow$ & \multicolumn{1}{c|}{SSIM$\uparrow$} &
            PSNR$\uparrow$ & \multicolumn{1}{c}{SSIM$\uparrow$}\\
            \hline
        \multicolumn{1}{l|}{CDDM} & \pmb{26.36} & \multicolumn{1}{c|}{\pmb{0.870}} & \pmb{29.12} & \multicolumn{1}{c|}{\pmb{0.861}} & \pmb{29.06} & \multicolumn{1}{c}{\pmb{0.861}} \\
        \multicolumn{1}{l|}{DDS~\cite{chung2023fast}} & 25.92 & \multicolumn{1}{c|}{0.862} & 28.80 & \multicolumn{1}{c|}{0.860} & 28.73 & \pmb{0.861} \\
        \multicolumn{1}{l|}{FBPConvNet~\cite{jin2017deep}} & 17.56 & \multicolumn{1}{c|}{0.317} & 17.28 & \multicolumn{1}{c|}{0.305} & 17.13 & 0.299 \\
        \multicolumn{1}{l|}{ADMM~\cite{boyd2011distributed}} & 18.71 & \multicolumn{1}{c|}{0.517} & 19.72 & \multicolumn{1}{c|}{0.504} & 19.50 & 0.500 \\
    \bottomrule
    \end{tabular}}
    \label{tab_res_walnuts}
\end{table}

\begin{table}[t!]
    \centering
    \caption{Quantitative results from different reconstruction algorithms for AAPM dataset.}
    \small
    \resizebox{\columnwidth}{!}{
    \begin{tabular}{lcccccc}
    \toprule
            \multicolumn{1}{c|}{} & \multicolumn{2}{c|}{Axial Plane} & \multicolumn{2}{c|}{Coronal Plane} & \multicolumn{2}{c}{Sagittal Plane}\\
        \multicolumn{1}{c|}{} & PSNR$\uparrow$ & \multicolumn{1}{c|}{SSIM$\uparrow$} & PSNR$\uparrow$ & \multicolumn{1}{c|}{SSIM$\uparrow$} &
            PSNR$\uparrow$ & \multicolumn{1}{c}{SSIM$\uparrow$}\\
            \hline
        \multicolumn{1}{l|}{CDDM} & \pmb{33.71} & \multicolumn{1}{c|}{\pmb{0.911}} & \pmb{35.22} & \multicolumn{1}{c|}{0.885} & \pmb{34.13} & 0.887 \\
        \multicolumn{1}{l|}{DDS~\cite{chung2023fast}} & 32.98 & \multicolumn{1}{c|}{0.893} & 34.71 & \multicolumn{1}{c|}{\pmb{0.902}} & 33.60 & \pmb{0.905} \\
        \multicolumn{1}{l|}{FBPConvNet~\cite{jin2017deep}} & 16.41 & \multicolumn{1}{c|}{0.532} & 18.45 & \multicolumn{1}{c|}{0.532} & 17.75 & 0.521 \\
        \multicolumn{1}{l|}{ADMM~\cite{boyd2011distributed}} & 20.08 & \multicolumn{1}{c|}{0.556} & 21.16 & \multicolumn{1}{c|}{0.511} & 19.82 & 0.500 \\
    \bottomrule
    \end{tabular}}
    \label{tab_res_aapm}
\end{table}

\subsection{Parametric Selection}

The selection of hyper-parameters is crucial in CT reconstruction and significantly impacts the performance characteristics of CDDM. Key factors such as the initial noisy image in pixel diffusion, the number of inference steps $T$, and the coefficient $\lambda_{max}$ are instrumental in determining image quality. Several experiments were made on the Walnut dataset.

\begin{figure}[!t]
    \centering
    \includegraphics[width=0.99\columnwidth]{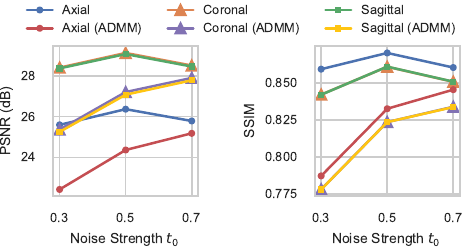}
    \vspace{-.1in}
    \caption{
    The relationship between noise strength $t_0$ added to the latent diffusion results and the final reconstruction images.} 
        \label{fig_walnut_noisestrength}
    \vspace{-.1in}
\end{figure}

The initial noisy image depends on the low-quality image $\mathbf{x}_0$ and the noise strength $t_0$. We introduce noise of varying strengths to the low-quality images generated by ADMM and latent diffusion to serve as the starting point for CDDM's reverse diffusion process in pixel space. The quantitative results are displayed in Fig.~\ref{fig_walnut_noisestrength}. Compared to images directly noised by ADMM, those generated through CDDM's whole process with a high-quality initial image achieve superior reconstruction quality, which is consistent with previous research that initializes the starting from a better initialization may achieve increased stability and performance~\cite{chung2022come}. For CT reconstructions initialized from latent diffusion, the optimal performance is achieved at $t_0=0.5$. When $t_0=0.3$, the noise perturbation on the low-quality image is insufficient, leading to fewer reconstruction steps and, consequently, sub-optimal results. Poorer performance at $t_0=0.7$ may be attributed to increased instability in the diffusion model during the initial phase of the reconstruction as strength increases~\cite{song2023solving}. This instability could be exacerbated by the DM, as it attempts to predict the image again while compensating for the discrepancy brought about by data consistency. For the pixel diffusion process beginning with ADMM, increasing the noise added to the image tends to improve CT reconstruction outcomes, primarily due to the inadequate reconstruction quality of ADMM images.

\begin{figure}[!t]
    \centering
    \includegraphics[width=0.99\columnwidth]{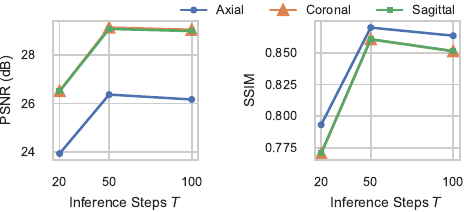}
    \vspace{-.1in}
    \caption{
    The relationship between total inference steps $T$ in the pixel diffusion process and the final reconstruction images.} 
        \label{fig_walnut_inferencesteps}
    \vspace{-.1in}
\end{figure}

The impact of inference steps $T$ is illustrated in Fig.~\ref{fig_walnut_inferencesteps}. Similar to the results with noise strength $t_0$, CDDM achieves its best performance at $T=50$. At $T=20$, the reverse process suffers from insufficient steps and relatively poorer convergence. The mediocre performance at $T=100$ can be attributed to the fact that an increase in steps does not enhance the refinement of the reconstructed image; instead, data consistency introduces additional errors.

\begin{figure}[!t]
    \centering
    \includegraphics[width=0.99\columnwidth]{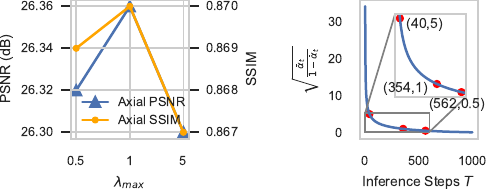}
    \vspace{-.1in}
    \caption{
    The effect of $\lambda_{max}$ on PSNR and SSIM on the axial plane and the relationship between $\sqrt {\frac{{{{\bar \alpha }_t}}}{{1 - {{\bar \alpha }_t}}}}$ and inference steps $T$. The right graph indicates the steps closest to $\sqrt {\frac{{{{\bar \alpha }_t}}}{{1 - {{\bar \alpha }_t}}}} =0.5,1,5$.} 
        \label{fig_lambda_max}
    \vspace{-.1in}
\end{figure}

The influence of the coefficient $\lambda_{max}$ on CT reconstruction is not significant, as shown in Fig.~\ref{fig_lambda_max}. The PSNR and SSIM on the axial plane remain essentially unchanged with variations in $\lambda_{max}$. Although the introduction of this parameter was initially intended to accommodate the convergence of model training, it suffices to assign a certain weight to the error term of data consistency in the objective, $\Delta \mathbf{x}_0 = {\Tilde{\mathbf{x}}_0^{recon}} - {\mathbf{x}_0}$, to allow the model to learn this discrepancy. Furthermore, as $t$ increases, $\sqrt {\frac{{{{\bar \alpha }_t}}}{{1 - {{\bar \alpha }t}}}}$ decreases at a slower rate. Selecting $\lambda{max}=0.5,1,5$ essentially covers the range of steps in the image reconstruction process.

\subsection{Ablation Studies}
We have proposed novel methods for sparse-view CT reconstruction based on diffusion generative models, conducting a series of ablation studies on the Walnut dataset. Our studies compared the impact of using DM on image quality within the CDDM method and investigated the effects of different data consistency approaches (gradients in $xyz$ versus $z$, specialized ADMM versus standard ADMM) on reconstruction accuracy, as summarized in Table~\ref{tab_ablation}.

\begin{table*}[t!]
    \centering
    \small
    \caption{Quantitative results of ablation studies on Walnut dataset.}
    \begin{tabular}{ccccccccc}
    \toprule
            \multicolumn{2}{c|}{ADMM} & \multirow{2}{*}{DM} & \multicolumn{2}{|c|}{Axial Plane} & \multicolumn{2}{c|}{Coronal Plane} & \multicolumn{2}{c}{Sagittal Plane}\\
            \multicolumn{1}{c}{Gradient Directions} & \multicolumn{1}{c|}{Type} &  &  \multicolumn{1}{|c}{PSNR$\uparrow$} & \multicolumn{1}{c|}{SSIM$\uparrow$} & PSNR$\uparrow$ & \multicolumn{1}{c|}{SSIM$\uparrow$} &
            PSNR$\uparrow$ & \multicolumn{1}{c}{SSIM$\uparrow$}\\
            \hline
            $xyz$ & \multicolumn{1}{c|}{Specialized} & \multicolumn{1}{c|}{\checkmark} & 26.36 & \multicolumn{1}{c|}{0.870} & 29.12 & \multicolumn{1}{c|}{0.861} & 29.06 & \multicolumn{1}{c}{0.861} \\
            $xyz$ & \multicolumn{1}{c|}{Specialized} & \multicolumn{1}{c|}{$\times$} & 25.40 & \multicolumn{1}{c|}{0.846} & 28.43 & \multicolumn{1}{c|}{0.833} & 28.37 & \multicolumn{1}{c}{0.833} \\
            $z$ & \multicolumn{1}{c|}{\text{-}} & \multicolumn{1}{c|}{\checkmark} & 25.72 & \multicolumn{1}{c|}{0.858} & 28.27 & \multicolumn{1}{c|}{0.843} & 28.23 & \multicolumn{1}{c}{0.844} \\
            $z$ & \multicolumn{1}{c|}{\text{-}} & \multicolumn{1}{c|}{$\times$} & 25.11 & \multicolumn{1}{c|}{0.852} & 27.96 & \multicolumn{1}{c|}{0.832} & 27.93 & \multicolumn{1}{c}{0.832} \\
            $xyz$ & \multicolumn{1}{c|}{Standard} & \multicolumn{1}{c|}{\checkmark} & 26.13 & \multicolumn{1}{c|}{0.870} & 28.74 & \multicolumn{1}{c|}{0.858} & 28.70 & \multicolumn{1}{c}{0.858} \\
    \bottomrule
    \end{tabular}
    \label{tab_ablation}
\end{table*}

The inclusion of DM was shown to significantly enhance image quality. Within the one-step reconstruction process, DM mitigates discrepancy caused by data consistency, aligning the data distribution more closely with the diffusion prior and thereby enhancing reconstruction results. When only $z$ directional gradients are considered in data consistency, DM yields certain improvements, though not markedly. This may be due to the neglect of $xy$ gradients, where ADMM fails to sufficiently smooth the image on the $xy$ plane. Consequently, applying DM on axial plane images does not effectively eliminate the training-sampling discrepancy, leading to less obvious effects.

Analyzing the consideration of $xy$ gradients, employing DM while accounting for gradients in all directions substantially improves image reconstruction quality. However, considering $xy$ gradients without using DM results in only slight quality improvements and may cause the image to become slightly blurry. This observation aligns with previous research findings that the usage of an $xy$ prior shifts the results somewhat away from the well-trained diffusion prior~\cite{chung2023solving}, highlighting the efficacy of DM in recovering the diffusion prior.

Comparing our specialized ADMM to standard ADMM reveals that both approaches yield similar SSIM metrics, yet specialized ADMM achieves significantly higher PSNR. This indicates its superior ability to delineate image boundaries, demonstrating stronger performance in solving the inverse problem. This effectiveness arises from its capability to decompose a complex problem into two moderately complex sub-problems, rather than pairing a complex problem with a simple one~\cite{ramdas2016fast, barbero2014modular}.

\section{Discussion and Conclusion}
Utilizing diffusion generative models to address the challenge of sparse-view CT reconstruction offers a solution for the inverse problem with an extremely limited number of projection views, where both iterative methods and end-to-end deep learning models struggle to produce high-quality images. Typically, diffusion generative models are employed to generate high-quality images with a difference from the true outcome, followed by integration with measurements for data consistency. This involves modifying the score or optimizing denoised images. However, this reconstruction approach still faces issues such as deviation from the diffusion prior distribution and long inference times. To address these challenges, we introduce the CDDM framework, which initially generates low-quality images using latent diffusion, then proceeds with reconstruction in pixel space starting from noisy images rather than Gaussian noise. For one-step reconstruction, we apply specialized ADMM which processes image gradients in each direction separately for data consistency and innovatively propose discrepancy mitigation to reduce deviation from the diffusion prior.

In the domain of diffusion generative models, the training-sampling discrepancy is seldom analyzed, with most efforts focused on addressing the exposure bias issue in the unconditional generation problems, i.e., the error accumulation impact on generation quality. In the process of solving inverse problems with diffusion models, the effect of data consistency on diffusion generation is more pronounced, as this guidance originates solely from subspace decomposition without considering the diffusion process's influence. Therefore, analyzing the relationship between the diffusion process and data consistency can facilitate the improvement of image quality. Our proposed DM, by analyzing the relationship between reconstructed images and original images after ensuring measurement consistency, introduces a novel training objective aimed at returning data to its original manifold. Experimental results demonstrate that DM significantly enhances imaging quality. This approach of guided diffusion is akin to per-step self-recurrence which aims to reduce guidance artifacts~\cite{lugmayr2022repaint, bansal2023universal}. However, unlike previous methods, DM utilizes specially trained models to more effectively minimize such errors and is particularly effective during the latter half of the reverse diffusion process, likely due to the increased difficulty in identifying artifacts in lower-quality images.

Another objective of diffusion generative models is to reduce the number of function evaluations (NFE). Computational costs within the latent space are significantly lower, with the computation time for latent diffusion with a downsampling factor of $f=8$ being approximately one-tenth that within pixel space for the same NFE. Starting the diffusion process in pixel space from a noisy image can halve the NFE ($t_0 =0.5$). However, since the DM process requires diffusion model calculations, the computational cost of CDDM under similar conditions is comparable to that of the DDS diffusion process, but with half the number of data consistency iterations. Overall, our proposed CDDM achieves comparable computational costs to conventional diffusion model-based sparse-view CT reconstruction methods while yielding superior reconstruction results.

For ill-posed inverse problems, regularization is commonly introduced to approximate a well-posed problem, ensuring unique and stable solutions. TV regularization, which assumes the data to be piecewise constant functions with reasonable edge sets, aligns with the inherent characteristics of images and is thus frequently employed in image reconstruction. ADMM, a widely used optimization method, has been applied extensively in solving linear inverse problems such as trend filtering and image denoising. For cases where $\mathbf{A} = \mathbf{I}$, some studies consider separating the gradients in TV for faster convergence. However, for image reconstruction problems where $\mathbf{A} \neq \mathbf{I}$, existing work typically solves the gradient term as a whole. We propose specialized ADMM, which separates gradients in different directions and decomposes a complex problem into two more manageable sub-problems, achieving higher reconstruction quality than standard ADMM. This method may also be extended to other types of linear inverse problems.

This paper focuses on analyzing the proposed CDDM framework for addressing the sparse-view CT reconstruction problem, thoroughly validating the effectiveness of the method. For other inverse problems solved using diffusion generative models, such as natural image inpainting and deblurring, limited-angle CT reconstruction, or under-sampled Magnetic Resonance Imaging (MRI) reconstruction challenges, applying the DM approach to mitigate reconstruction errors may likewise prove effective. CDDM is the first to verify the efficacy of the two-stage diffusion model for sparse-view CT imaging, effectively leveraging the characteristics of different diffusion models to accelerate the image generation process, though it is not the first one to employ a series of diffusion models for medical imaging issues. This base-refiner image generation pattern can also be extended to other image generation contexts.

The main limitation of the proposed CDDM is the increased computational cost due to the DM process, which requires an additional diffusion process to mitigate the training-sampling discrepancy. It could be beneficial to explore the combination of the DM process and the denoised diffusion process to reduce the computational effort. Future work could also consider the reconstruction completely within the latent space, though it would cause issues of VAE such as image compression loss and many-to-one mapping~\cite{rout2023solving, song2023solving}, which are not as critical for natural image inverse problems but are crucial for high-precision medical image reconstruction.

In summary, we design a cascaded diffusion model with discrepancy mitigation for sparse-view CT reconstruction. The two-stage structure improves the efficiency by rationally utilizing the latent and pixel diffusion models. The discrepancy mitigation process aims to reduce the training-sampling discrepancy induced by data consistency. The specialized ADMM method for data consistency is designed based on the features of the 3D CT reconstruction problem. The proposed CDDM outperforms common diffusion reconstruction models in terms of image quality and computational cost, providing new insights for similar inverse problems.

{
    \small
    \bibliographystyle{ieeenat_fullname}
    \bibliography{main}
}

\end{document}